\documentclass[showpacs,preprintnumbers,amsmath,amssymb]{revtex4}
\usepackage{graphicx}
\input epsf
\usepackage{dcolumn}
\usepackage{bm}

\begin{document}
\title{The linearization method and new classes of exact solutions in cosmology}
\author{A.V. Yurov}
\email{artyom_yurov@mail.ru}
\affiliation{I. Kant Russian State University, Theoretical Physics
Department, Al.Nevsky St. 14, Kaliningrad 236041, Russia}

\author{A.V. Astashenok}
\email{artyom.art@gmail.com}
\affiliation{I. Kant Russian State
University, Theoretical Physics Department, Al.Nevsky St. 14,
Kaliningrad 236041, Russia}

\begin{abstract}
We develop a method for constructing exact cosmological solutions
of the Einstein equations based on representing them as a
second-order linear differential equation. In particular, the
method allows using an arbitrary known solution to construct a
more general solution parameterized by a set of 3\textit{N}
constants, where \textit{N} is an arbitrary natural number. The
large number of free parameters may prove useful for constructing
a theoretical model that agrees satisfactorily with the results of
astronomical observations. Cosmological solutions on the
Randall-Sundrum brane have similar properties. We show that
three-parameter solutions in the general case already exhibit
inflationary regimes. In contrast to previously studied
two-parameter solutions, these three-parameter solutions can
describe an exit from inflation without a fine tuning of the
parameters and also several consecutive inflationary regimes.
\end{abstract}

 \maketitle

\medskip
\section{Introduction}
A method for constructing and analyzing exact cosmological
solutions of the Einstein equations based on representing them as
a second-order linear equation (we call this the linearization
method in what follows) was presented in \cite{Chervon} (other
methods for constructing exact solutions in cosmology can be found
in \cite{Barrow}-\cite{Barrow2}). Indeed, it is easy to see that
in the case of the of the flat Friedmann metric, the third power
of the scale factor $\psi =a^3$ satisfies the equation:
\begin{equation}
\frac{d^2\psi}{dt^2}=\frac{9}{2}\left(\rho-p\right)\psi,
\label{a3}
\end{equation}
where $\rho$ is the density and $p$ is the pressure of the matter
filling the universe. Here and hereafter, we use the system of
units with $8\pi G/3=c=1$. In the case where a minimally coupled
scalar field $\phi$ with the self-interaction potential $V(\phi)$
is dominant and in the presence of a cosmological constant with
the density $\Lambda$ Eq. (\ref{a3}) formally coincides with the
Schr\"{o}dinger equation
\begin{equation}
\frac{d^2\psi}{dt^2}=(U-\lambda)\psi, \label{Schr}
\end{equation}
where the potential is $U(t)=9V$, and the spectral parameter is
$\lambda=-9\Lambda$. In (\ref{Schr}), the quantity $V$ is assumed
to be a function of time: $V(t)=V(\phi(t))$, which was called the
history of the potential in \cite{Chervon}. Giving an explicit
form $U(t)$ together with the corresponding boundary conditions
allows finding the general solution of (\ref{Schr}). An important
consequence of this investigation is that the regime is
independent of or weakly dependent on the type of the potential,
which is quite significant for the whole theory. Unfortunately,
the problem of the end of inflation turns out to be substantially
more difficult, and solving it in the framework of the approach
described above apparently involves additional assumptions (the
authors of \cite{Chervon} proposed modifying the potentials to
make them depend on the temperature. The establishment of a
Friedmann regime can then be described as a phase transition in
the matter of the early universe). The study of Eq. (\ref{Schr})
in its application to cosmology was continued in \cite{Yurov1} è
\cite{Yurov2}, where the Darboux transformation was used to
construct new exact solutions (a similar technique was used in
\cite{Yurov3} to construct exact solutions on the brane and on the
encompassing space carrying an orbifold structure).

Here, we present a modification of this method. The Einstein
equations imply that the function $\psi_{n}=a^{n}$, where $n$ n is
an arbitrary (not necessarily integer) number, satisfies a
Schr\"{o}dinger equation with a function $U_{n}$, that is a linear
combination of the density and pressure. If we assume that the
universe is filled with a minimally coupled scalar field, then
$U_{n}$ is a linear combination of the potential $V(\phi)$ and the
kinetic term, and it is therefore no longer reasonable to call
$U_{n}$ the history of the potential. Hereafter, we call the
quantity $U_n$, the "potential" in quotation marks to distinguish
it from the self-interaction potential $V$.

Fixing a "potential" we can find solutions for the function
$\psi_{n}$ and thus find the scale factor $a_n=\psi_n^{1/n}$ as a
function of time. In general, a solution of the Schr\"{o}dinger
equation has the form
\begin{equation}
\psi_n(t)=c_1\psi_1(t)+c_2\psi_2(t), \label{supepoz}
\end{equation}
i.e., it depends on two arbitrary constants. The scale factor in
turn depends on three parameters: $a=a(n,c_1,c_2;t)$. This
circumstance allows constructing multiparameter solutions of the
Einstein equations as follows. We assume that the quantity $a(t)$
is determined, for example, by astronomical observation. Raising
this function to the $n_1$th power, we obtain a function denoted
by $\psi_1$ in (\ref{supepoz}). In the next step, we find $\psi_2$
based on the condition of linear independence, which we write as
$$
\psi_1\frac{d\psi_2}{dt}-\psi_2\frac{d\psi_2}{dt}=1,
$$
after which we find a three-parameter solution
$a=a(n_1,c^{(1)}_1,c^{(1)}_2;t)$. We can now repeat this procedure
and find a six-parameter solution
$$
a=a(n_1,c^{(1)}_1,c^{(1)}_2;n_2,c^{(2)}_1,c^{(2)}_2;t)
$$
and so on. After $N$ steps, we obtain a solution that depends on
$3N$ parameters. On the other hand, knowing the scale factor in
the Friedmann cosmology allows computing all other
characteristics, such as the Hubble parameter, the acceleration,
the density, and the pressure. The photometric distance, which is
a crucial quantity for testing models, is a function of six
parameters; therefore, using the above procedure sufficiently many
times, we can make the model consistent with observations.

Unfortunately, the effectiveness of this method is considerably
reduced when considering Friedmann models with a nonzero spatial
curvature. In this case, the function $\psi_{n}=a^{n}$ for an
arbitrary $n$ n satisfies not the Schr\"{o}dinger equation but an
equation with an additional nonlinear term. There are no
meaningful techniques for integrating such an equation; it is
therefore generally difficult to find a solution depending on two
arbitrary constants, which is necessary for a complete examination
of the problem, similar to the investigation in\cite{Chervon}. It
would be interesting to show that the conclusions in
\cite{Chervon} also hold for $k=\pm 1$. Fortunately, solving the
complicated nonlinear equation discussed above is unnecessary for
this: it suffices to set $n=1$ instead of $n=3$. In this case, the
wave function is the scale factor itself. One of the two
cosmological Einstein equations in the Friedmann metric is a
second-order linear equation with a potential proportional (with a
minus sign) to $\rho+3p/c^2$. The physical meaning of the
"potential" is also more transparent. If we set the cosmological
term to zero, then if the potential is negative, then the strong
energy condition is satisfied, and if it is positive, then this
condition is violated, which generally implies inflation.
Therefore, the problem of studying inflationary regimes becomes
much simpler, even when compared with flat ($k=0$) models.

This paper is organized as follows. In Sec. 2, we formulate the
linearization method exactly, i.e., the reduction of the Friedmann
equations to a Schr\"{o}dinger equation for arbitrary $n$.
Moreover, we prove a similar assertion for the cosmology on the
Randall{Sundrum I brane (RS-I). In Sec. 3, we give several
examples of exact solutions. We see that three-parameter families
of solutions are much richer in properties than the two-parameter
solutions studied in \cite{Chervon}. In particular, such
"potentials" can lead to solutions describing several inflationary
stages. Perhaps our universe is currently undergoing one of them.

\section{The linearization method}

We consider the Einstein equations in the Friedmann metric:
\begin{eqnarray}
  \frac{\dot{a}^2}{a^{2}} &=& \rho-\frac{k}{a^{2}},
  \label{Ein1}\\
  \frac{\ddot{a}}{a} &=& -\frac{1}{2}\left(\rho+3p\right) \label{Ein2}.
\end{eqnarray}
We assume that $a=a(t)$, $p=p(t)$, $\rho=\rho(t)$ is a solution of
these equations for $k=0$. Then the function $\psi_{n}=a^{n}$ is a
solution of the Schr\"{o}dinger equation
\begin{equation}
{\ddot \psi_{n}}=U_{n}(t)\psi_{n}, \label{2}
\end{equation}
where the "potential" is
\begin{equation}
U_{n}(t)=n^{2}\rho-\frac{3n}{2}(\rho+p). \label{Un}
\end{equation}

If the universe is filled with a scalar field $\phi$ with the
Lagrangian $L=\frac{\dot{\phi}^{2}}{2}-V(\phi)$, then
\begin{equation}
U_{n}=\frac{n(n-3)}{2}\dot{\phi}^{2}+n^{2}V(\phi). \label{Un-1}
\end{equation}

{\bf Remark 1}. For $n=3$, the "potential" is $U_{3}=9V(\phi(t))$.
This case was studied in detail in \cite{Chervon}, where this
quantity, as already noted, was called the history of the
potential because $U_3$ appears in the equations as a function of
time ($U=U(t)$), not a function of the field variable $\phi$.
Nevertheless, the potential $U\sim V$, the physical meaning of the
potential $U$ therefore seems clear. If $n\ne 3$, then $U_n$ is a
certain linear combination of the kinetic term ${\dot\phi}^2/2$
and the self-interaction potential $V$ (ñì. (\ref{Un}),
(\ref{Un-1})), and the physical meaning of such a $U$ is not so
evident.

Nevertheless, the effectiveness of the method presented in
\cite{Chervon} (and developed in \cite{Yurov1},
 \cite{Yurov2}), precisely consists
in reducing a complex nonlinear problem to a linear equation. This
allows ¯nding full two-parameter solutions, which exhibit
inflationary behavior under very general assumptions. The fact
that $U$ basically coincides with $V$ was not used anywhere in
those papers and therefore played no role. Similarly, we here
consider a generalization of this method to arbitrary $n$.

Furthermore, the physical meaning of the "potential" $U_3$ is
clear only for a universe filled with a scalar field. If we
consider a universe in which, for example, electromagnetic
radiation is dominant, the physical meaning of the quantity $U_3$
becomes ambiguous.

{\bf Remark 2}. If we assume that the universe contains a nonzero
vacuum energy with the density $\rho_{_\Lambda}c^2$, in addition
to the matter fields, then Eq. (\ref{2}) takes the form of the
spectral problem
\begin{equation}\label{Scr-11}
 \ddot{\psi}_{n}=(U_{n}(t)-\lambda_{n})\psi_{n},
\end{equation}
where the spectral parameter is
$\lambda_{n}=-3n^{2}\rho_{_\Lambda}/3$. Just as for Eq.
(\ref{Schr}), we can consider a problem for the eigenvalues and
the eigenfunctions of Eq. (\ref{Scr-11}) if we specify homogeneous
initial conditions. As noted in \cite{Chervon}, Eq. (\ref{Schr})
has the form of a quantum mechanical problem with a discrete
spectrum. The fact that each such solution only admits a bounded
or countable set of allowed values of the cosmological constant
(if we specify homogeneous initial conditions) may clarify the
question of the actual value of the cosmological constant.

We note that if a solution of (\ref{Scr-11}) is known, then we can
use (\ref{Ein1}) and (\ref{Scr-11}) to find the scalar field and
the potential:
\begin{equation}
\phi(t)=\pm\frac{\sqrt{2}}{\sqrt{3n}}\int
dt\sqrt{\frac{\dot{\psi}_{n}^{2}}{\psi_{n}^{2}}-U_{n}+\lambda_{n}},\label{phit}
\end{equation}
\begin{equation}
V(t)=\frac{1}{3}\left(\frac{U_{n}}{n}+\frac{3-n}{n^{2}}\left(\frac{\dot{\psi}_{n}^{2}}{\psi_{n}^{2}}+\lambda_{n}\right)
\right). \label{Vt}
\end{equation}
We can obtain the dependence $V=V(\phi)$ from these expressions,
although it is clearly not always possible to do this explicitly.
In the general case, a solution of (\ref{2}) has the form
\begin{equation}
\Psi_{n}=c_{1}\psi_{n}+c_{2}\hat{\psi}_{n},\label{general}
\end{equation}
where $\hat{\psi}_{n}$ is a linearly independent solution with the
same potential:
\begin{equation}
{\hat\psi}_n(t)=\psi_n(t)\int^t \frac{dt'}{\psi_n^2(t')}\equiv
\psi_n(t)\xi(t). \label{xi}
\end{equation}
Equation (\ref{general}) allows proving the following assertion.
\newline
{\bf Assertion}. Let $a=a(t)$ be a solution of (\ref{Ein1}) and
(\ref{Ein2}) for $k=0$ and the corresponding $\rho$ and $p$. Then
the three-parameter function $a_n=a(t;c_1,c_2,n)$ of the form
\begin{equation}
a_n=a\left(c_1+c_2\int\frac{dt}{a^{2n}}\right)^{1/n}, \label{an}
\end{equation}
is a solution of (\ref{Ein1} and \ref{Ein2}) for a new energy
density $\rho_n$ and pressure $p_n$ satisfying the condition
\begin{equation}
n^2\rho_n-\frac{3n}{2}\left(\rho_n+p_{n}/c^{2}\right)=
n^2\rho-\frac{3n}{2}\left(\rho+p/c^{2}\right). \label{inv}
\end{equation}

{\bf Remark 3}. In general, this assertion holds for $k=0$. If
$k=\pm 1$, then it holds only if $n=0,1$.

{\bf Remark 4}. A similar assertion can be formulated for
solutions describing the RS-I brane. In this case, the Friedmann
system is modified by taking the brane tension $\sigma$ into
account and becomes
\begin{equation}
\begin{array}{cc}
\displaystyle{\left(\frac{\dot{a}}{a}\right)^2=\rho
\left(1+\frac{\rho}{2 \sigma}\right),}
\\
\\
\displaystyle{-2\frac{\ddot{a}}{a}=\rho+3p+\frac{\rho}{\sigma}\left(2
\rho+3p \right),}
\\
\\
\end{array}
\label{novobran}
\end{equation}
It is easy to see that the function
 $\psi_n\equiv a^n$ satisfies the linear Schr\"{o}dinger equation
\begin{equation}
\frac{{\ddot \psi_n}}{\psi_n}=W_n, \label{Scrone}
\end{equation}
with the potential
\begin{equation}
\begin{array}{l}
\displaystyle{ W_n=\frac{n}{2}\left(2n \rho-3\left(\rho+p
\right)+\frac{\rho}{\lambda}\left(n
\rho-3\left(\rho+p\right)\right)\right)=}\\
\\
\displaystyle{=\frac{n}{2}\left[2n(K+V)-6K+\frac{1}{\lambda}(K+V)\left(n(K+V)-6K\right)\right]},
\end{array}
 \label{Wn}
\end{equation}
where $V=V(\phi)$ and $K={\dot\phi}^2/2$. Hence, the method of
generating $3N$-parameter families of solutions described above
can also be applied to cosmology on a brane. An example of an
exact two-parameter solution (with $n=3$) was given in \cite{AAV},
 where a linearization method for a simple anisotropic cosmological
model was described.

In what follows, we consider several examples of exact solutions
based on an integrable potential of the Schr\"{o}dinger operator,
following \cite{Chervon}.

\section{Generating exact solutions with a given $U_{n}(t)$}

We consider the model "potentials"
$$
U_{n}(t)=\mu^{2}t^{2},\eqno{(A)}
$$
$$
U_{n}(t)=-\frac{2\lambda_{0}}{\cosh^{2}(\lambda_{0}t)}.\eqno{(B)}
$$
We study the solutions for potentials (A) and (B) for the possible
existence of inflationary regimes and exit from inflation.

{\bf Potential (A)}. For potential (A), the solution of Eq.
(\ref{Scr-11}) with zero boundary conditions as
$t\rightarrow\pm\infty$ has the form
$$
\psi_{n}=A^{n}H_{s}(\mu t)\exp(-\mu t^{2}/2),
$$
where $A$ is a constant and $H_{s}(\mu t)$ are the Hermite
polynomials of order $s$. The corresponding evolution of the scale
factor is
\begin{equation}
a(t)=AH_{s}^{1/n}(\mu t)\exp(-\mu t^{2}/2n).
\label{votono}
\end{equation}
In the simplest case, we have $\psi_{n}=\exp(-\mu t^{2}/2)$. In
this case, the dependen on $n$ can be eliminated by a simple
rede¯nition of the parameter $\mu$. This solution was considered
in \cite{Chervon}, and we therefore do not consider it further.

Choosing the wave function of the ground state as the solution of
problem (\ref{Scr-11}) is not obligatory. We can consider
solutions in $L^2$ that correspond to excited levels, but these
solutions cannot be used on the entire interval on which they are
defined. The point is that according to the oscillation theorem,
the wave function of the $s$th excited level has $s$ zeros, each
of which in the cosmological context corresponds to a singularity
with the scale factor tending to zero if $n>0$, and to in¯nity (a
Big Rip singularity) if $n<0$. We can use solutions for excited
levels with numbers $s>1$ and consider the dynamics described by a
part of the eigenfunction on an interval. Such a universe begins
and/or ends its existence at the corresponding singularity. For
example, if we take the function $\psi_{n}\sim H_{1}(\mu
t)\exp(-\mu t^2/2)$ as a solution of (\ref{Scr-11}) with the
"potential" $U_{n}=\mu^{2}t^{2}$ and let $n$ be -1, then the
evolution of the scale factor in this universe can be written as
\begin{equation}
a(t)=\frac{a_{0}t_{0}}{t}{\rm
e}^{\mu(t^{2}-t_{0}^{2})/2}.\label{aBigRip}
\end{equation}
This solution describes a universe in which the scale factor is
equal to in¯nity at $t=0$ (a Big Rip singularity), takes its
minimum value at $t=\mu^{-1/2}$ and then begins a never-ending
inflationary stage. Using formulas (\ref{phit} and \ref{Vt}), we
can also find the asymptotic behavior of the scalar field and the
potential for $t\sim 0$ and as $t\rightarrow \infty$. The kinetic
term of the energy density of the scalar field is negative, i.e.,
the evolution described by (\ref{aBigRip}) corresponds to a
phantom field ñîîòâåòñòâóåò ôàíòîìíîìó ïîëþ (see, e.g.,
\cite{phantom2}-\cite{Odin} for phantom fields). Equation
(\ref{aBigRip}) corresponds to a negative cosmological constant
$\Lambda=-3\mu$.

{\bf Potential (B)}. The solution for potential (B) for
$\lambda_{n}=-\lambda^{2}\leq 0$ (which corresponds to a
nonnegative value of the cosmological constant) has the form
\begin{equation} \label{Solch}
\psi_{n}=c_{1}(\lambda-\lambda_{0}\tanh(\lambda_{0}t)) {\rm
e}^{\lambda t}+c_{2}(\lambda+\lambda_{0}\tanh(\lambda_{0}t)){\rm
e}^{-\lambda t}.
\end{equation}

If $\lambda=\lambda_{0}$, then formula (\ref{Solch}) simplifies
considerably:
$$
\psi_{n}=\frac{C}{\cosh(\lambda_{0}t)}.
$$
For positive $n$, the evolution of the scale factor corresponding
to the function $\psi_{n}$ describes with an inflationary regime
on the interval $(-\infty,t_{0})$, where $t_{0}$ is the inflection
point of the function $\cosh^{1/n}(\lambda_{0}t)$.

An interesting class of solutions can also be obtained in the
simplest case $\lambda_{n}=0$. Then (\ref{Solch}) becomes
\begin{equation}
\psi_{n}=C\tanh(\lambda_{0}t).
\end{equation}
If $n\geq1$, then the solution for the scale factor describes a
universe leaving the singular state at the moment $t=0$ and
asymptotically approaching a stationary state as
$t\rightarrow\infty$ with $\ddot{a}<0$ during the entire
evolution. If $0<n<1$, then the universe undergoes an in°ationary
phase until a certain instant and then asymptotically approaches a
stationary state.

In the case where $\lambda>\lambda_{0}$, $c_{1}>0$ and $c_{2}=0$,
there is a solution
$$
\psi_{n}=c_{1}(\lambda-\lambda_{0}\tanh(\lambda_{0}t)){\rm
e}^{\lambda t},
$$
\begin{equation} \label{doubleinfl}
a=c_{1}^{1/n}(\lambda-\lambda_{0}\tanh(\lambda_{0}t))^{1/n}{\rm
e}^{\lambda t/n}.
\end{equation}
This solution corresponds to evolution without singularities and
is interesting because for certain values of the parameters
$\lambda_{0}$, $\lambda$ and $n$, formulas (\ref{doubleinfl})
describe a universe that undergoes inflationary expansion on some
interval $(0,t_{1})$, then inflation ends, and a secondary
inflationary period begins at a time $t_{2}>t_{1}$.To determine
when such a situation occurs, we consider $\ddot{a}$ at the
initial instant. Using (\ref{doubleinfl}), we see that
$$
\ddot{a}(0)\sim (1-n)y^{4}-2y^2+1,
$$
where we introduce the notation $y=\lambda_{0}/\lambda<1$.
Therefore, the second derivative of the scale factor is
nonnegative at the initial instant if
$$
0<y^{2}\leq y_{0}^{2}=\frac{1-\sqrt{n}}{1-n}.
$$
A further investigation shows that satisfaction of this condition
means that the universe immediately enters an inflationary regime
if $n\geq 1$. If $y^{2}>y_{0}^{2}$, then solution
(\ref{doubleinfl}) describes an evolution with an inflationary
regime starting at a time $t_{0}>0$. If $n<1$ then in the narrow
interval $y_{0}^{2}-\Delta<y^2<y_{0}^{2}$, where $\Delta\ll
y_{0}^{2}$, there is a period of noninflationary expansion, after
which the universe again enters a period of accelerating
expansion. For example, if $n=0.25$ and $\lambda_{0}=0.8\lambda$,
then the universe undergoes accelerating expansion in the interval
$(0,0.08/\lambda)$, followed by a noninflationary expansion stage
of length approximately $0.4/\lambda$. For $t>5/\lambda$, the
scale factor can be considered to change according the exponential
law $a\sim {\rm e}^{4\lambda t}$.

We conclude by considering the asymptotic behavior of the
potential of the scalar field during the early stage of the
evolution of such a universe. We limit ourself to terms linear in
time in the decomposition. Using (\ref{phit}) and (\ref{Vt}), we
find that for $t\ll1/\lambda,$
$$
\phi(t)\approx\phi_{0}\pm\sqrt{\frac{2}{3n}}y^{2}\lambda t,
$$
$$
U(t)\approx\frac{\lambda^{2}}{3}\left(\frac{3-n}{n^{2}}y^{4}-\frac{6}{n^{2}}y^{2}-
\frac{2(3-n)}{n^{2}}y^{4}\left(1-y^{2}\right)\lambda t\right).
$$
Therefore, if there is an initial inflation, then it corresponds
to a slow-roll regime. The potential slowly decreases as the
scalar ¯eld increases or decreases linearly.

\section{Conclusion}

We have proposed a relatively simple method for generating exact
solutions of the Einstein-Friedmann equations. We used
mathematical transformations to reduce the problem to solving the
Schr\"{o}dinger equation for the function $a^{n}$ with a
"potential" proportional to $n^{2}\rho-3n(\rho+p/c^2)/2$. We
focused on studying the inflationary regime and the exit from it.

Three-parameter families of solutions exhibit a much richer
repertoire of behaviors than do the two-parameter solutions
studied in \cite{Chervon}. Nevertheless, these solutions, like the
solutions described in \cite{Chervon}, have inflationary phases
under quite general assumptions. This is an indication that
inflation is not something exotic found only in a limited number
of models. On the contrary, an inflationary regime seems a fairly
common occurrence in cosmology not requiring any special initial
assumptions.

Unlike two-parameter solutions, three-parameter solutions can have
several consecutive inflationary phases, i.e., they can not only
describe inflation but also describe an exit from the inflationary
phase without a special fine tuning of the parameters. Moreover,
the existence of such solutions can be seen as indirect evidence
for the existence of a {\em unified} realistic model that contains
not only an inflationary phase (with an exit from it without a
fine tuning) in the early universe stage but also a later
inflation, which our own universe is perhaps undergoing currently.

Finally, the procedure for constructing a three-parameter solution
can be repeated an arbitrary number $N$ of times, yielding
$3N$-parameter solutions. The presence of a sufficiently large
number of free parameters can be considered a defect of a theory
but, on the contrary, can be useful for fitting the theoretical
model to observational data. We plan to return to this question in
future investigations. Hence, the described method may prove quite
fruitful in cosmology.

\medskip

{\bf Acknowledgments}. The authors thank S. V. Chervon for the
useful discussions of the obtained results. The authors are also
grateful to the anonymous referee for the valuable comments on the
main text of the article. This work was supported in part by the
Russian Foundation for Basic Research (Grant No. 08-02- 91307-IND
a).

\end{document}